\documentclass[sigchi]{acmart}
\AtBeginDocument{%
  }

\setcopyright{none}
\renewcommand\footnotetextcopyrightpermission[1]{}
\begin{document}

\title{Why Algorithms Remain Unjust: Power Structures Surrounding Algorithmic Activity}

\author{Andrew Balch}
\email{xxv2zh@virginia.edu}
\orcid{0009-0005-1282-7480}
\affiliation{%
  \institution{University of Virginia}
  \city{Charlottesville}
  \state{Virginia}
  \country{USA}
}

\renewcommand{\shortauthors}{Balch}

\begin{abstract}
  Algorithms are unavoidable in our social lives, yet often perpetuate social injustices. The popular means of addressing this is through algorithmic reformism: fine-tuning algorithms themselves to be more fair, accountable, and transparent. However, reformism fails to curtail algorithmic injustice because it ignores the power structure surrounding algorithms. Heeding calls from critical algorithm studies, I employ a framework developed by Erik Olin Wright to examine the configuration of power surrounding algorithmic systems in society (Algorithmic Activity). Algorithmic Activity is unjust because it is dominated by economic power. To create socially just Algorithmic Activity, the power configuration must instead empower end users. I explore Wright's symbiotic, interstitial, and raptural transformations in the context of just Algorithmic Activity. My vision for social justice in algorithmic systems requires a continuous (re)evaluation of how power can be transformed in light of current structure, social theories, evolving methodologies, and one’s relationship to power itself.  
\end{abstract}



\received{5 December 2024}

\maketitle

\pagestyle{plain}

\section{Introduction}

With generative artificial intelligence (AI) squarely at the forefront of public consciousness, it seems more pertinent now than ever before to discuss the role of algorithms in society, how they perpetuate injustice, and most importantly, why all ameliorative efforts have arguably failed. It is no longer deniable that algorithms shape the ways we see and act upon the social world, often encoding existing systems of inequality \cite{kordzadeh_algorithmic_2022, eubanks_automating_2018-1, birhane_algorithmic_2021, eidelson_patterned_2021, burrell_society_2021, devlin_power_2023}. The overwhelming response from the industry (researchers, developers, and organizations) to these glaring issues has been characterized by algorithmic reformism. \citet{polack_beyond_2020} summarizes algorithmic reformism as the methods ``whereby calculated evaluations and critiques of algorithm logic motivate its redesign without changing the underlying problems and design requirements it is supposed to satisfy". Algorithmic reformism discussions center the algorithm itself, asking how its development and use can be more ethical, fair, and human-centered in the hopes of relieving algorithmic bias \cite{kaur_trustworthy_2022, barocas_fairness_2023}. In practice, algorithmic reformism usually involves setting ethical development and use guidelines, attempting to explain decisions, establishing human oversight, assessing algorithm bias with mathematical metrics, and adjusting outcomes to more evenly distribute an algorithm's impact across social groups. 

The reformist approach has been widely criticized: guidelines are soft and frequently undermined, transparency is limited by technological advancement and proprietary designs, human-centered design often ignores implicit biases and norms, and distributive justice-based fairness is local in scope \cite{kasirzadeh_algorithmic_2022, maas_machine_2023, green_good_2019, resseguier_power_2023, polack_beyond_2020, davis_algorithmic_2021}. At the root of all these criticisms is the recognition that traditional algorithmic reformism ignores the larger power structures and social dynamics that surround algorithms themselves. It is evident that algorithmic reformism, while important, has been largely unsuccessful. Critical algorithm studies have arisen in response to the persistence of algorithmic injustice, pointing out the failings of reform-only approaches and arguing for perspectives that look around the algorithm, taking structural injustices and systems of power into account \cite{kasirzadeh_algorithmic_2022, maas_machine_2023, green_good_2019, resseguier_power_2023, polack_beyond_2020, davis_algorithmic_2021, kasy_fairness_2021, lee_how_2019, joyce_toward_2021, eidelson_patterned_2021, hoffman_five_2022, walker_algorithmic_2023, hoffmann_where_2019}. My aim in this paper is to diagnose the root cause of these algorithmic injustices by providing an analysis of the power configuration surrounding algorithms in society. In doing so, I hope to promote a reflexive consideration of one's relationship to these power structures, emphasizing existing tools and theories that may be used for socially-just transformations.

My analysis herein relies on Erik Wright's discussion of real utopias \cite{wright_transforming_2013}. In ``Transforming Capitalism through Real Utopias", Wright argues that “many forms of human suffering and many deficits in human flourishing are the result of existing institutions and social structures” and that, by transforming these structures, we can reduce suffering and approach flourishing \cite{wright_transforming_2013}. This transformation starts by identifying ``moral principles for judging social institutions", using them to criticize existing institutions, exploring viable alternatives, and finally, proposing transformations ``for realizing those alternatives" \cite{wright_transforming_2013}. Wright conducts this process in the context of Economic Activity: the exchange of goods and services. I focus on employing the tools Wright provides to judge what I call Algorithmic Activity, diagnose the power structure that perpetuates unjust Algorithmic Activity, and explore transformative solutions. By doing so, our understanding of how algorithms reproduce injustice can be deepened.

Other works have applied frameworks rooted in social theory to analyze the relationship between power structures and algorithmic systems. For example, \citet{miceli_data-production_2022} extends Foucault's concept of a dispositif to data-work that has been outsourced to Latin America, exploring how social structures maintain normative exercises of power that propagate through the act of outsourcing. A framework of power from Cristiano Castelfranchi has been applied to understand how end-users of machine learning systems are systematically dominated by both developers and users of these systems \citet{maas_machine_2023}. \citet{polcumpally_artificial_2022} assume a global perspective, using systems theory and a triple helix model to analyze how nation-states are vying for power over AI. In adopting Wright's framework, I wish to take a whole-systems approach to analyzing algorithmic activity that is not focused on a single perspective, dynamic, or mechanism. This is also driven by the fact that Wright does not just provide a holistic view of how (and why) things are the way they are. Crucially, Wright also gives us an idea of \textit{how things ought to be}. It is this tension between how things are and how they ought to be that allows for the exploration of \textit{transformations}: methods that can fundamentally alter the normative configuration of power such that it more closely resembles that of a utopia. 

The structure of my paper is as follows: First, I define Algorithmic Activity and demonstrate how it does not meet Wright's standards for social justice: equality, democracy, and sustainability. Second, I formulate the current power configuration surrounding Algorithmic Activity, positioning it in contrast with Wright's socially-just utopia. In doing so, I argue that Algorithmic Activity is unjust because the power structure that surrounds it is dominated by economic power (as opposed to state or, in a utopia, social power). Third, I summarize Wright's transformative strategies for social empowerment and place them in the context of Algorithmic Activity. Last, I walk through the different roles these strategies may play within a hypothetical research project. 

\section{Algorithmic Activity, Society, and Injustice}

I define Algorithmic Activity as \textit{the ways in which algorithms are researched, developed, trained, and deployed within society}. In this way, Algorithmic Activity looks beyond the impact of individual algorithms, the sole concern of algorithmic reformism. Building on calls from critical algorithm theorists, this definition considers the actors that shape Algorithmic Activity itself, and thereby encompasses the powers surrounding the existence and behavior of algorithms. 
Furthermore, I define the term ``algorithm" as any \textit{automatic} process by which an input is \textit{systematically} and \textit{intentionally} mapped to an output.
Normative ideals about proximity, relation, and the realm of possible end results are built into an algorithm and act as pre-defined constraints upon the algorithm itself, as well as any interaction with it \cite{beer_social_2017, resseguier_power_2023, panagia_possibilities_2021}. In other words, the developers of an algorithm hold implicit social norms and biases that are imbued into the algorithm. These norms and biases ultimately shape the algorithm's output and, coincidentally, the individual lives in which their decisions play a role.

The system surrounding Algorithmic Activity, as we know it now, does not fulfill Wright's moral principles of equality, democracy, or sustainability \cite{wright_transforming_2013}. Algorithmic Activity is unequal in the sense that it perpetuates biases against vulnerable social groups \cite{obermeyer_dissecting_2019, noble_algorithms_2018, adam_write_2022, wolfe_american_2022, abid_persistent_2021, eubanks_automating_2018-1}. It is undemocratic in the sense that those impacted by algorithmic decisions often know nothing about the algorithm itself, or even that it exists \cite{maas_machine_2023, pasquale_black_2015}. Lastly, it is unsustainable in the sense that algorithms, as they are, do not ensure equal or greater access to social, economic, and environmental resources for future generations \cite{van_wynsberghe_sustainable_2021, green_good_2019, galaz_artificial_2021, coleman_ais_2023, patterson_carbon_2021, li_making_2023}. 

When an algorithm is deployed in society, it has the capacity to both \textit{automatically} shape the social world (through its results and decisions), and to be shaped by it (through data and normative design). 
Therefore, Algorithmic Activity may be better understood as a uniquely modern social institution. In sociology, an institution provides a normative framework for moving through the world. Institutions are self-reproducing, creating a constant feedback loop where institutions shape individuals and vice versa. 
So, what makes Algorithmic Activity distinct from traditional institutions such as language, the family, the state, a school, or a corporation? The answer is threefold: opacity, efficiency, and minimal human intervention. The opaque ``black box" is a popular characterization of an algorithm, communicating the inability of the general public to understand its inner workings. The opacity of Algorithmic Activity does not stop ``under the hood" of the algorithm itself. Algorithmic Activity play roles in key life decisions each and every day, often without our knowledge \cite{pasquale_black_2015, lustig_algorithmic_2016}. The efficiency of Algorithmic Activity is much more obvious, so much so that it is an algorithm's main selling point. Encoded, systematic processes are quicker to formulate and execute than the more ambiguous nature of other institutions, which are characterized by slow transformations in the social consciousness or the rational crawl of the bureaucratic process. However, this efficiency comes at a great social, economic, and environmental cost. The social cost is perhaps the most concerning: algorithms are enabled to act back on society with minimal human intervention. No matter how opaque a bureaucratic institution may be, there is always an individual moving the cogs of the machine and facilitating its impact on society. When algorithms are deployed to inform or make decisions, there is often no human actively influencing the ways in which the algorithm produces its results. While an algorithm's designers may claim a responsible, human-centered development approach, this does not guarantee a human-centered implementation in the real world \cite{green_good_2019}. 

It is evident that Algorithmic Activity does not just violate the moral principles of equality, democracy, and sustainability. It perpetuates these injustices in ways that are more invisible, efficient, and fundamentally disconnected from humanity than ever before seen. To have any hope in transforming this pattern, we must analyze the social power configuration that surrounds and maintains Algorithmic Activity, emphasizing how it does not currently align with 

\section{Configuration of Power in Algorithmic Activity}

In his paper, \citet{wright_transforming_2013} provides a visual vocabulary to conceptualize how the three types of power (State, Economic, and Social) interact and exert control over economic activity. \textit{State power} is rooted in the creation and enforcement of rules, \textit{economic power} is rooted in ``the use of economic resources", and \textit{social power} is ``rooted in the capacity to mobilize people for cooperative, voluntary collective actions" \cite{wright_transforming_2013}. Wright argues that it is not as simple as any one power having complete control over economic activity. Instead, the three powers coexist in a hybrid configuration that can be more, or less, capitalist, statist, or socialist. There is a push and pull between each power that defines how control is exerted. Furthermore, I add that an individual in society is not constrained only to the exercise of social power.  Rather, individual actors exert multiple types of power by virtue of the role they play in Algorithmic Activity. This is to say that each person influences and is influenced by the power structures surrounding Algorithmic Activity.

\begin{figure}[t]
    \centering
    \includegraphics[width=0.55\linewidth]{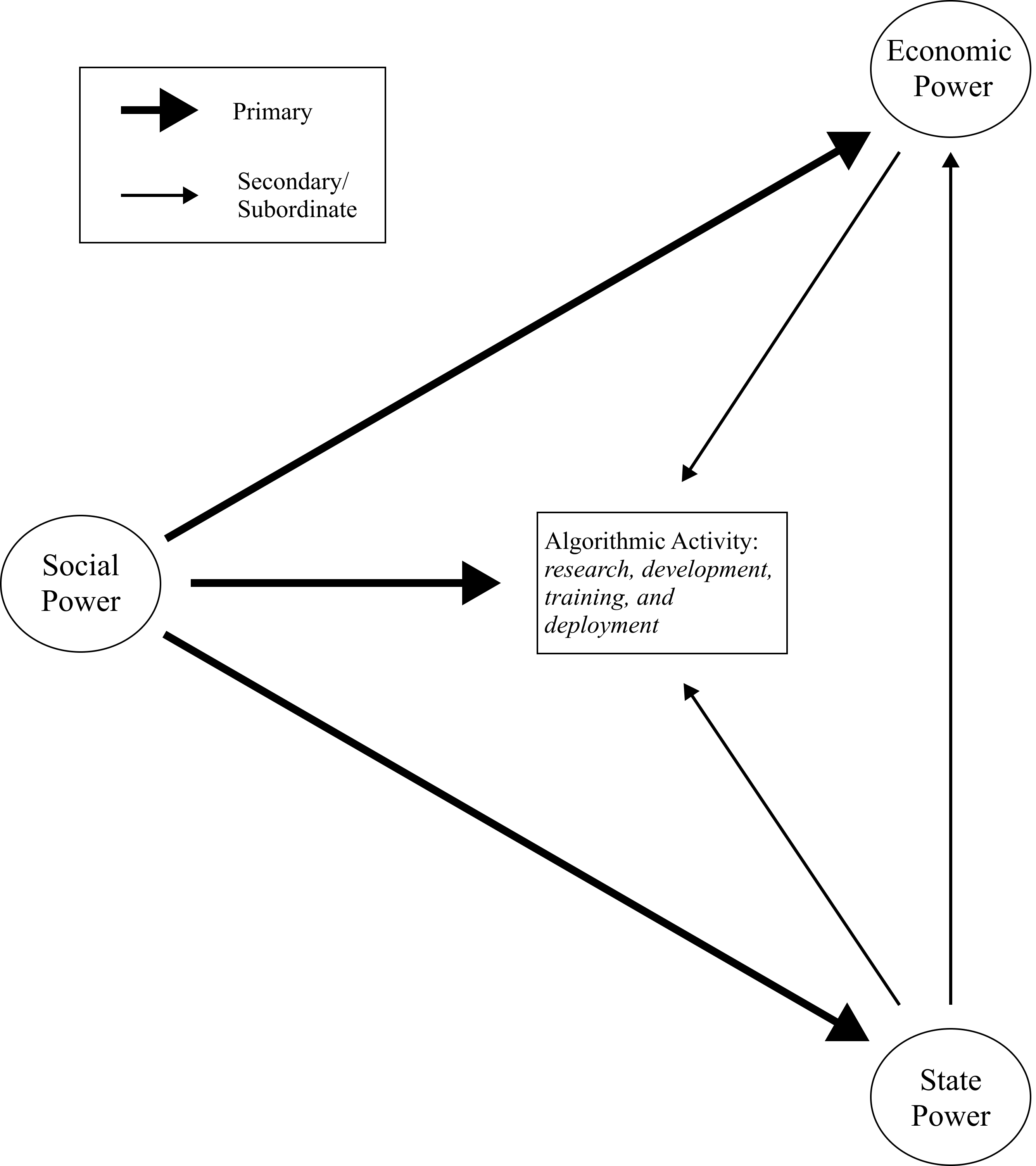}
    \caption{A utopic power configuration where Social Power dominates Algorithmic Activity, and all other powers are subordinated to it.}
    \label{fig:soc-power}
\end{figure}

For reference, Figure \ref{fig:soc-power} represents Wright's configuration of social empowerment \cite{wright_transforming_2013}. 
The interaction between different forms of power, and ultimate control over Algorithmic Activity, is shown by the arrows. An arrow's boldness indicates the strength and autonomy of the power exerted along its path, where bolder is primary and thinner is secondary or subordinate to a dominant power. 
This power configuration promotes Wright's principles of a just society (a utopia) because Algorithmic Activity would be ``controlled through the exercise of social power", ensuring that neither economic nor state power impede the best interests of society \cite{wright_transforming_2013}. If one is to be consciously engaged in the task of furthering social justice with respect to Algorithmic Activity, then they must first ask how (and why) the structure of power surrounding this activity does not align with the principles of social justice.

\begin{figure}[t]
    \centering
    \includegraphics[width=0.6\linewidth]{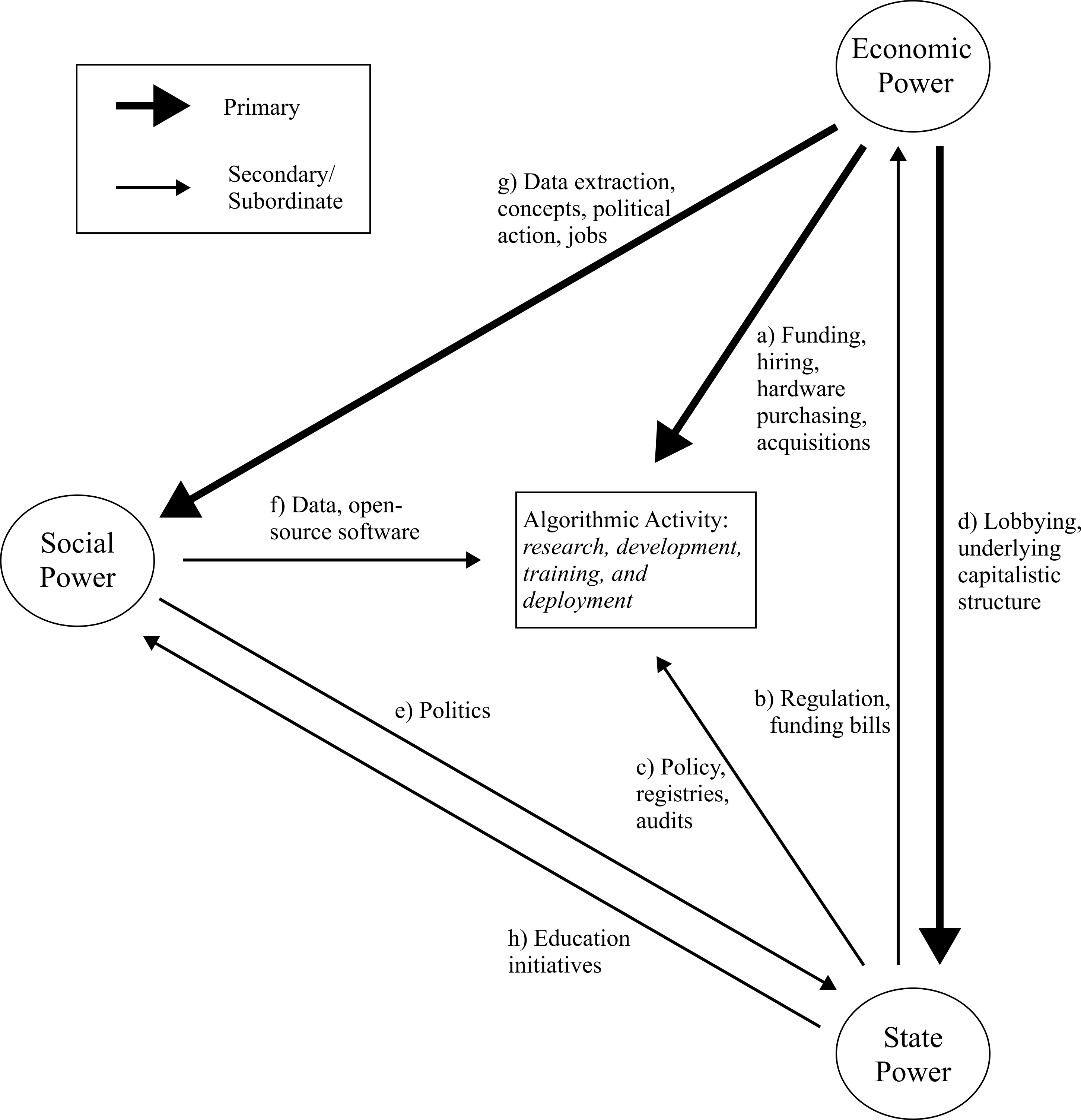}
    \caption{The modern configuration of power around Algorithmic Activity is one of Economic Empowerment.}
    \label{fig:algo-power}
\end{figure}

Figure \ref{fig:algo-power} is the result of my analysis of the configuration of power surrounding Algorithmic Activity. This diagram demonstrates just how far we are from socially just Algorithmic Activity. Economic Empowerment is the dominant force in this power configuration, exerting control over not only Algorithmic Activity, but social and state power as well. As a means of validating this power structure, I will describe how each form of power controls Algorithmic Activity, illustrating how the influence of other powers shapes the nature of this control. 

\subsection{Economic Power (fig. \ref{fig:algo-power}a)}

It is evident that economic actors with the most resources, such as Big Tech (Alphabet/Google, Amazon, Apple, Meta/Facebook, and Microsoft), can fund bigger research projects, hire the best people, own the most powerful hardware, and acquire any smaller company or project. In other words, the means of algorithmic production is effectively owned by Big Tech, which affords them ultimate control over Algorithmic Activity, especially in academia \cite{abdalla_grey_2021, hoffman_managing_2017}. Even the means of \textit{responsible} production is owned by Big Tech. It has been shown that the standards for responsible machine learning (ML) development, largely influenced by Big Tech, are difficult for less cash-rich organizations to meet, leaving any progress in this area subject to the whim of Big Tech \cite{hopkins_machine_2021}. 

\subsubsection{State Power upon Economic Power (fig. \ref{fig:algo-power}b)} Only state power has a clear, direct influence in the exercise of economic power over Algorithmic Activity. State actors can regulate the economy and pass bills that provide funding for projects. The state may even directly intervene to break up economic power that has become too dominant (as in a monopoly). However, this has yet to occur in practice within the context of Algorithmic Activity.

\subsection{State Power (fig. \ref{fig:algo-power}c)}

State actors have recently expressed an increased interest in Algorithmic Activity. There is the potential for the state to step in and set policies surrounding the development and use of more powerful algorithms, mandate the creation of registries for models and training datasets, and conduct audits of sensitive systems. The European Union (EU) has been at the forefront with its Digital Services Act (DSA), which aims to provide protections via ``algorithmic transparency and accountability", and its AI Act, which establishes ``obligations for providers and users depending on the level of risk from artificial intelligence" \cite{european_parliament_eu_2023, european_centre_for_algorithmic_transparency_european_2024}. 
The full impact of these legislative initiatives have yet to be seen as the DSA just recently came into effect in February 2024, and the AI act was adopted in March 2024 and is yet to be fully applicable.
Data protection regulations like the EU's General Data Protection Regulation (GDPR) are also an important exercise of state power.

\subsubsection{Economic Power upon State Power (fig. \ref{fig:algo-power}d)} The primary reason for the strength of this interaction is the underlying structure of the capitalist state in countries such as the United States (U.S.) \cite{wright_transforming_2013}. Because the state depends on the capitalist economic substructure, the dynamic in Figure \ref{fig:algo-power}c is, in some ways, subordinate to economic power. This manifests through actions such as lobbying for government grants, looser regulations, or more relaxed policies surrounding Algorithmic Activity. 

\subsubsection{Social Power upon State Power (fig. \ref{fig:algo-power}e)} Politics is the central means by which social power mediates state power in a democratic society \cite{wright_transforming_2013}. People participate in politics and form parties to shape the behavior of the state. 

\subsection{Social Power (fig. \ref{fig:algo-power}f)}

Social power's control over Algorithmic Activity originates from the data generated by society as well as the development of free and open-source software. The data we create by moving through the world is the main avenue by which social power influences the design of algorithms. Algorithms are explicitly (as in ML) or implicitly (as in sorting methods) designed around data, so they consequently encode various social norms, values, concepts, and relations that are latent in the data itself \cite{beer_social_2017, panagia_possibilities_2021, resseguier_power_2023}. In the age of large language models and vision language models, even art itself is appropriated for its data. 
Another social activity that directly furthers Algorithmic Activity is the creation of free and open-source software. In theory, open-source software is a form of nonhierarchical cooperative Algorithmic Activity where individual social actors organize to create algorithms that are not directly profit-motivated. 

\subsubsection{Economic Power upon Social Power (fig. \ref{fig:algo-power}g)} Social power's control over Algorithmic Activity is subordinate to economic power. As in Figure \ref{fig:algo-power}e, this is because the underlying substructure of U.S. society is capitalism and economic power owns the means of algorithmic production (fig. \ref{fig:algo-power}a). More precisely, economic power controls the exercise of social power through conventional means such as the labor market and political action (e.g. uncapped donations to super PACs). This occurs alongside more algorithm-specific methods, such as extracting social data \cite{zuboff_age_2023, couldry_data_2019} and shaping the social concept of the algorithm. The sale of the algorithm as a product is the mechanism that allows these avenues of subordination. Economic actors sell us products that we use to navigate the world around us, and that then extract our data as we do so (e.g. smartphones, social media, ChatGPT). This directly undermines the social control over Algorithmic Activity through data (fig. \ref{fig:algo-power}f) for two main reasons. First, we are not always in control or aware of the data we generate \cite{pasquale_black_2015}. Second, there is an implicit bias surrounding whose data is collected, who collects it, and who determines its ``truth". Access to technology is stratified across socioeconomic status \cite{yu_algorithmic_2020, lutz_digital_2019}, data collection is conducted primarily by Western universities and corporations, and ground-truth labels are determined by a relatively small portion of society. Thus, this relatively small portion of society has a disproportionate influence on algorithms that impact society at large. Control exerted by free and open-source software (fig.  \ref{fig:algo-power}f) is also facilitated by economic power. In practice, the most prominent software libraries for developing ML algorithms (e.g. PyTorch, TensorFlow, and CUDA) are maintained or otherwise funded by Big Tech (Meta/Facebook, Google, and Nvidia, respectively). Even when freely given this software, the means of producing the most powerful, state-of-the-art algorithms (e.g. GPT-4) remain far out of reach for individual social actors \cite{yu_algorithmic_2020}. Proprietary hardware (Nvidia's CUDA-capable GPUs) is necessary to train algorithms like neural networks at-scale or sometimes just as one-offs. This leads the individual interested in contributing to Algorithmic Activity back to Big Tech. There, they can purchase computing capability à la carte via platforms like Amazon Web Services or Google Colab. The last and most potently invisible force exerted by economic power is the ability to influence the social concept of the algorithm. Since economic power is the dominant force in Algorithmic Activity, economic actors like Big Tech can shape the algorithm as a concept through its design, use, and surrounding marketing. Through these mechanisms, AI has become a buzzword associated with productive efficiency and rational, objective results, despite the fact that the reality of Algorithmic Activity is anything but fact-driven \cite{beer_social_2017}.

\subsubsection{State Power upon Social Power (fig. \ref{fig:algo-power}h)} State power has the potential to shape the social understanding of Algorithmic Activity through public education initiatives, such as those proposed by a recent Virginia executive order \cite{governor_of_virginia_governor_2024}. Recently, state powers have taken action to outright ban social media for users under the age of 16 in Australia \cite{ortutay_australia_2024}. For better or for worse, this legislation directly impacts the exercise of social power with respect to Algorithmic Activity for this social group.
\\

It should now be evident how the power configuration surrounding Algorithmic Activity in Western, capitalist nations
such as the U.S. serves to empower the interests of economic actors (fig. \ref{fig:algo-power}) at the expense of the interests of society (fig. \ref{fig:soc-power}). Algorithmic Activity remains unjust because it is a fundamentally profit-driven activity. Economic power is interested in a just society so much as such a society rewards those economic actors with more capital than the unjust status-quo. Now that we have reflected upon the tension between socially-just Algorithmic Activity and the nature of Algorithmic Activity as we know it today, we must consider how the power configuration can be transformed to more closely resemble one where social power is dominant. 

\section{Transforming Power for a More Just Society}

Wright describes three main strategies for transformation through which an alternative power structure can be realized \cite{wright_transforming_2013}. \textit{Symbiotic transformations} are those that work to extend and deepen ``institutional forms of social empowerment", often through the state \cite{wright_transforming_2013}. \textit{Interstitial transformations} seek out the edges of the existing system in which to ``build new forms of social empowerment" \cite{wright_transforming_2013}. Lastly, \textit{raptural transformations} deconstruct the existing power structure and reform it around ``new, emancipatory institutions" \cite{wright_transforming_2013}. These three transformative strategies are weak on their own, and more effective in combination. 
Here, I provide an overview of existing methodologies for conducting Algorithmic Activity that would be socially-empowering, organized according to Wright's three transformations. 
Keeping with the theme of reflexivity, my goal in this section is to prompt a thoughtful engagement with existing transformative strategies, which can be extended to one's unique role within the power structure surrounding Algorithmic Activity.

\subsection{Symbiotic Transformations}

Algorithmic reformism constitutes symbiotic transformations of Algorithmic Activity. I have already demonstrated that the overall weakness of reformist initiatives is how they place the algorithm itself front and center, ignoring the surrounding power configuration and associated systemic injustices. Simply put, reformism attempts to deepen social empowerment without threatening the status-quo, and does so under the control of the dominant power. 
Take the most potent tool in the reformist's belt, algorithmic transparency, as an example. On the surface, making an algorithm's decisions 'transparent' involves making its strengths and weaknesses evident to all stakeholders \cite{kaur_trustworthy_2022}. When achieved, it can improve stakeholder trust, empowering them with a more comprehensive understanding about the model and how it acts back upon them. Unfortunately, economic actors benefit from the opacity of the ``black box". Transparency is in direct opposition to the interests of the economic actor, where the protection of intellectual property (IP) is of the upmost importance. Therefore, the inner-workings of the most powerful, influential algorithms (e.g. GPT-4) are often kept secret to protect IP \cite{pasquale_black_2015}. But this secrecy provides a dual function. When an algorithm inevitably reproduces social biases, companies, governments, and other organizations can hide behind the notion of the ``rogue algorithm" and claim a lack of understanding in its processes. By doing so, the owners of the algorithm can shirk responsibility for its actions. The notion of the algorithmic ``black box" is thus a great asset to economic power, and they can use their dominance over Algorithmic Activity to cyclically create bigger, more opaque algorithms that have to be ``explained".
Besides its local scope and reliance on Big Tech, reformism is a failure because there is a lack of institutional forms of social empowerment to entrench. The state must first fully implement enforceable policies and regulations upon Algorithmic Activity before symbiotic transformations can have any real impact. Beyond the pending EU regulations, audit requirements for algorithmic decision support systems have been discussed, but these would need to be carefully designed to promote social justice and would likely require state force to be effective \cite{vecchione_algorithmic_2021}. Similarly, avenues for recourse against algorithmic decisions can be provided \cite{alkhatib_street-level_2019}. This recourse may use other symbiotic ideals of fairness, accountability, and transparency to identify a wrongdoing, then pursue action against it in the form of an appeal or compensation. Again, effective recourse that empowers oppressed social groups relies on the goodwill of economic power as well as state oversight. The state may also reach for the carrot instead of the stick, opting to pass funding bills that incentivize economic actors to participate in Algorithmic Activity that is socially just. Symbiotic transformations are important endeavors, but they are not enough to make for more just Algorithmic Activity on their own. 

\subsection{Interstitial Transformations}

As the weaknesses surrounding algorithmic reformism are made more apparent and critical algorithm studies gain traction, proposals for interstitial transformations have begun to surface. Some strategies use algorithms themselves to analyze the structural biases in Algorithmic Activity \cite{kuhlberg_advancing_2020}, others target data (fig. \ref{fig:algo-power}g) attempting to de-Westernize it and ``recognize nonmainstream ways of knowing the world through data" \cite{milan_big_2019}. Alternatives to algorithmic reformism have been proposed that situate reformist approaches in the context of the broader social problems they are trying to address, evaluating them against non-technical social justice reforms \cite{kordzadeh_algorithmic_2022, polack_beyond_2020, mohamed_decolonial_2020}. Participatory or democratic design seeks to actively involve communities and social groups in the development of algorithms that would impact them \cite{maas_machine_2023, bondi_envisioning_2021, birhane_power_2022}. Going a step further, \citet{harrington_denizen_2024} formulates an activist-led design which is led by the community. Confronting ``ethics washing" in AI, the setting of ethics guidelines that fails to account for power structures, Resseguier \cite{resseguier_power_2023} calls for a ``power to" ethics that emphasizes the ``opening of possibilities" for social groups that have historically been dominated and discriminated against. While it is not specific to Algorithmic Activity, collective bargaining may also be an effective interstitial transformation, creating the missing interaction between social power and economic power (fig. \ref{fig:algo-power}). Each of these methods would create new avenues for social empowerment by working within the power structure itself. 

\subsection{Raptural Transformations}

A raptural transformation in Algorithmic Activity could be as dramatic as breaking up the data and compute monopoly that is Big Tech or redistributing technological resources. More nuanced reforms have also been suggested. \citet{joyce_toward_2021} sees the discipline of sociology take a leading role in shaping algorithms. This envisioned future is rooted in the ability for sociologists to ``identify how inequalities are embedded in all aspects of society and to point toward avenues for structural social change" \cite{joyce_toward_2021}. In a similar vein, \citet{green_good_2019} argues that computer science often takes a technology-centric, ``greedy approach" to social reform, often resulting in long term harm. They call upon computer scientists to interrogate their normative understanding of ``social good", look to social thinkers and activists for more viable ways to bring about positive social change, and to actively work against the assumption that an accurate algorithm is the best solution to a problem. \citet{lazer_computational_2020} explores, among other things, how the university as an institution must be reorganized to emphasize interdisciplinary collaboration and effective ethical guidelines to meet the social challenges posed by the modern algorithm. Intersectional thought and practice has also inspired ``algorithmic reparation", where the implicit bias of an algorithm is leveraged to empower marginalized experiences \cite{davis_algorithmic_2021}. 

\section{Transformations in a Research Context}

To solidify these ideas, let us explore a hypothetical scenario in Algorithmic Activity. My goal as we walk through this scenario is to demonstrate how one's role and relationship to power can be reflexively (re)evaluated as well as how the aforementioned transformations may be applied in practice to yield a more socially-empowered Algorithmic Activity. In choosing the scenario, I wished to deviate from more popular discussions about recidivism prediction or credit scoring to show how the ideas presented in this paper can be applied to projects that are less controversial, or perhaps even altruistically-motivated. 
With this said, consider a team of researchers that wish to create an algorithm that determines how food should be distributed from a food bank. This algorithm has a clear impact on society, it determines how much food a person gets from the food bank. The goal for our researchers is an \textit{optimal} algorithm where the algorithm's decisions result in the most food given to the most people. Thus, they have a fairly normative research question: Is the proposed algorithm more or less ``optimal" than other algorithms? Let us assume that the starting point for our researchers is an algorithm that manages to distribute all the day's food to the most people, but does so by giving out the least amount of food possible to everyone except for the last patron, who is given the remainder of the food. Such a solution is clearly ineffective at helping people, and must be transformed to be more egalitarian.

\subsection{Symbiotic Transformation}
So, our researchers turn to symbiotic transformations to fix this problem. Their research question must be adjusted to match this new strategy: Is the proposed algorithm more or less fair, accountable, and transparent than other algorithms? Now, the system is adjusted such that it distributes food equitably to everyone, explains why a patron received the amount they got, and offers them a fair avenue to dispute this amount. As a result, individuals are treated more fairly and even have more say in how this Algorithmic Activity impacts them. However, there is still room for more social empowerment. 

\subsection{Interstitial Transformation}
Applying an interstitial transformation requires looking beyond the algorithm itself to the social problem it is trying to address. In this scenario, the social issue is undernutrition. Identifying the root social problem allows our researchers to improve how it is addressed by their algorithm in a few ways: First, they can engage with social theories about undernutrition and use them to inform potential solutions. Second, they can directly involve patrons of the food bank in the development and use of their algorithm. Third, they can explore how food banks (and their algorithm, by association) under- or over-serve different social groups. Fourth, and most importantly, our researchers can ask whether their algorithm is better at combatting undernutrition than alternative avenues of food distribution. This research question is distinct because of how it focuses on the social problem. It is therefore much more valuable than normative or symbiotic questions alone because its answer guides future work in the direction that best addresses the social issue. This entire process is an exercise in social empowerment because it directly confronts a social issue, engages those affected, and diligently works to alleviate its impact through the most effective mechanisms available.
Note that interstitial transformations do not completely exclude the algorithm as a possible tool for fixing undernutrition, they simply ask what alternative approaches could be more effective. Any of these alternatives may also be able to use algorithms as a tool to increase their potency. 

\subsection{Raptural Transformation}
Sometimes, a social injustice is so persistent that the only solution is to depart from the structure that upholds it and forge new mechanisms of confrontation. In our scenario, undernutrition in disenfranchised social groups may be perpetuated by how funding and food donations are distributed to food banks themselves, by the material inequalities that determine access to healthy food, or by larger institutions like industrial farming, which may prioritize sheer magnitude over nutritional value. Power structures can even impede symbiotic and interstitial transformations. For example, our team of researchers may be rushed to complete this particular project by some deadline set by a funding agency, a publisher, or the organization they are employed by. In so doing, an upper limit is placed upon their capacity to consider symbiotic and interstitial transformations that would deepen social empowerment. 
In such a situation, it is of the upmost importance for our team of researchers to follow the lead of social theory. Social theorists are better equipped to deconstruct the power structure surrounding injustice and offer a viable alternative that is socially-just at a fundamental level. 
While it may be possible to use the algorithm as a tool for this deconstruction, such as to analyze and highlight disparities in food access in connection with larger structural forces, this should take a back seat to initiatives for genuine social change. And so, our research question becomes the same as the social theorist's: How are injustices embedded within society and what structural changes are necessary to make the world more just?

\section{Conclusion}

The persistence of algorithmic injustice within society indicates a deeper issue with how its prevalence has been addressed through algorithmic reformism. Inspired by Wright and proponents of critical algorithm studies, I offered a diagnosis of the modern power configuration surrounding Algorithmic Activity and demonstrated how economic power dominates the realm. To create socially just Algorithmic Activity, this configuration of economic empowerment must be transformed to empower social actors instead. I described how this could be achieved through different symbiotic, interstitial, and raptural transformations to Algorithmic Activity and provided an example of these transformations in practice. 

My vision for Algorithmic Activity that is equitable, democratic, and sustainable has two criteria. First, Algorithmic Activity must be fundamentally socially-driven. This means that algorithms are applied as tools when they can create a more just society, and set aside when they can not. Second, algorithmic problems must be seen as social problems. For example, we must recognize that bias perpetuated by an algorithm's decisions may not be able to be fully addressed by adjusting the algorithm. Sometimes the underlying social bias must be confronted in a deep and meaningful way first. Such an alternative requires Wright’s symbiotic, interstitial, and raptural transformations together. As a society, we need meaningful legislative protections against unjust Algorithmic Activity that uses reformist approaches for recourse against economic power. We need the literacy to ask if Algorithmic Activity actually addresses social issues, and the space within the current structure to ask these questions. We need to break away from being sold greedy, technocratic solutions and produce our own reparative alternatives, grounded in social theory. This all involves learning from the failure of lone algorithmic reformism and working to build new tools and mechanisms for social empowerment. Future work that strives for socially just Algorithmic Activity should implement the proposed transformations where they are appropriate, innovate new strategies for deepening social empowerment, and continuously re-evaluate what it means to be socially just in light of evolving power structures, algorithmic systems/methodologies, and social theories. 

\begin{acks}
I would like to thank Professor Afsaneh Doryab, Professor Andreja Siliunas, and Natalie Bretton for their help and support developing this paper.
\end{acks}

\bibliographystyle{ACM-Reference-Format}
\bibliography{main}

\end{document}